\documentclass[aps,preprint,groupedaddress,showpacs]{revtex4}

\usepackage{graphicx}
\usepackage[fleqn]{amsmath}
\usepackage{amsfonts}
\usepackage{amssymb}
\usepackage{psfrag}

\newcommand{\Tens}[1]{\boldsymbol{\rm #1}}
\newcommand{\Vect}[1]{\boldsymbol{\rm #1}}
\newcommand{\Trans}[1]{{#1}^{\text{t}}}
\newcommand{\Det}[1]{\lvert #1 \rvert}

\DeclareMathOperator{\erf}{erf}
\newcommand{\ceil}[1]{\lceil #1 \rceil}
\begin{document}

\bibliographystyle{apsrev}


\title{Phase separation of a multiple occupancy lattice gas}

\author{Reimar Finken}
\email{rf227@cam.ac.uk}
\author{Jean-Pierre Hansen}
\author{Ard Louis}
\affiliation{Department of Chemistry\\
University of Cambridge\\
Cambridge CB2 1EW (UK)}

\date{\today}

\begin{abstract}
  A binary lattice gas model that allows for multiple occupancy of
  lattice sites, inspired by recent coarse-grained descriptions of
  solutions of interacting polymers, is investigated by combining the
  steepest descent approximation with an exploration of the
  multidimensional energy landscape, and by Gibbs ensemble Monte Carlo
  simulations. The one-component version of the model, involving on
  site and nearest neighbour interactions, is shown to exhibit
  microphase separation into two sub-lattices with different mean
  occupation numbers. The symmetric two-component version of the
  multiple occupancy lattice gas is shown to exhibit a demixing
  transition into two phases above a critical mean occupation number.

\end{abstract}
\pacs{61.20.Ja,64.60.-i,64.60.Cn,05.50.+q}

\maketitle


\section{Introduction}
\label{sec:introduction}

Simple fluids are dominated by excluded volume effects, which, within
lattice gas models, are accounted for by the single occupancy
constraint, whereby each site on a lattice can be occupied by at most
one molecule. Effective interactions between macromolecules or
self-assembled aggregates in complex fluids, on the other hand, can be
``soft'', i.e.\ lack an impenetrable core. A good example is the
effective pair potential between the centres of mass (CM) of
interacting polymer coils, obtained by taking statistical averages
over monomer conformations for fixed distances $r$ between the CMs
\cite{ref:1,ref:2}. Recent extensive simulations of self-avoiding walk
polymers carried out over a wide range of concentrations show that the
repulsive state-dependent effective CM pair potential is of roughly
Gaussian shape, of width governed by the polymer radius of gyration,
and of maximum amplitude $v(r=0) \approx k_B T$ \cite{ref:3}; this
behaviour reflects the fractal nature of polymers in good solvent,
leading to a low entropic cost at full overlap. Several other complex
fluids have been shown to exhibit ultrasoft coarse-grained
interactions, e.\ g.\ star polymers \cite{Likos98} or effective
particles considered within dissipative particle dynamics \cite{groot97}.

The penetrability of the corresponding ``Gaussian core'' (GC) model,
$v(r) = \epsilon \exp(-r^2)$, leads to interesting phase behaviour at
low temperatures ($T^* = k_B T / \epsilon \ll 1$), \cite{ref:4,ref:5},
but under conditions relevant for polymer solutions ($T^* \approx 1$),
the model behaves like a ``mean field'' fluid \cite{ref:6}. However
binary Gaussian core mixtures, characterised by different energy
scales $\epsilon_{\alpha\beta}$ for the three types of pairs, lead to
 phase separation for moderate couplings
\cite{ref:6,ref:7,ref:8}. This demixing, which occurs for purely
repulsive, penetrable interactions, is of a very different nature than
the usual phase separation of incompatible fluids, which is generally
driven by the long-range attractive intermolecular forces.

In this paper we examine the simplest lattice gas representation of
penetrable particles, in an effort to gain further insight into this
novel class of phase transitions. The penetrable nature of the
effective interaction is reflected by allowing \underline{multiple
  occupancy} of each lattice site. The simplest version of the model
involves only on-site interactions of particles of the two species,
with different energy penalties for the different types of pairs.
While a steepest descent estimate of the grand partition function
predicts phase separation, it will be shown that a more accurate
treatment reveals the spurious nature of this transition, as expected
for an effectively zero-dimensional system. The addition of nearest
neighbour interactions leads to a genuine demixing transition, similar
to that predicted for continuous versions of soft core mixtures
examined earlier within the random phase approximation (RPA)
\cite{ref:6,ref:7,ref:8}. We explore this behaviour by mean field
theories and Gibbs ensemble Monte Carlo simulations.  We also study
the topology of the energy landscape.

\section{The multi-occupancy model}
\label{sec:multi-occup-model}

Consider a $d$-dimensional lattice of $L$ sites and coordination
number $q$, which can accommodate particles of two species. The
occupancy of each site is characterised by the two-component vector of
integer occupation numbers $\Vect{n}_i = (n_1^{(i)},n_2^{(i)}),\, 1
\leq i \leq L$. Particles interact only when they are on the same site
or on nearest neighbour (n.n.) sites. The on-site and off-site
couplings are characterised by $2\times 2$ matrices of interaction
energies $\epsilon_{\alpha\beta}$ and $\eta_{\alpha\beta}\,(1\leq
\alpha,\beta \leq 2)$. In terms of the occupation numbers, the energy
of the system reads:
\begin{equation}
  \label{eq:1}
  E(\{\Vect{n}_i\}) = \frac{1}{2} \sum_i \Trans{\Vect{n}_i} \cdot
  \Tens{\epsilon} \cdot \Vect{n}_i - \frac{1}{2} \sum_i \sum_\alpha
  \epsilon_{\alpha\alpha} n_\alpha^{(i)}
  + \sum_{\langle ij \rangle} \Trans{\Vect{n}_i}\cdot \Tens{\eta}\cdot
  \Vect{n}_j,
\end{equation}
where the last summation is over n.n.\ sites only. For given chemical
potentials $\Vect{\mu} = (\mu_1,\mu_2)$ of the two species, the grand
partition function is:
\begin{multline}
  \label{eq:2}
  \Xi = \sum_{\{\Vect{n}_i\}} \frac{\exp\{\beta \sum_i
    \Vect{\mu}\cdot \Vect{n}_i\}}{\prod_{\alpha,i} n_\alpha^{(i)}!}
  \exp\{-\beta E(\{\Vect{n}_i\})\}\\
  = \sum_{\{\Vect{n}_i\}} \prod_{i=1}^L \frac{e^{\beta
      \Vect{\mu}^*\cdot \Vect{n}_i}}{n_1^{(i)}!n_2^{(i)}!}
  \exp\left\{-\beta \left[\frac{1}{2} \sum_i \Trans{\Vect{n}_i}
      \cdot \Tens{\epsilon} \cdot \Vect{n}_i + \sum_{\langle ij \rangle}
      \Trans{\Vect{n}_i}\cdot \Tens{\eta} \cdot \Vect{n}_j \right]\right\}.
\end{multline}
The first sum is over all possible distributions of occupation numbers
of the $L$ sites, and the effective chemical potentials $\mu_\alpha^*
= \mu_\alpha - \epsilon_{\alpha\alpha}/2$, $\Vect{\mu}^* =
(\mu_1^*,\mu_2^*)$ have been introduced.

The main task of this paper is to approximately evaluate the grand
partition function \eqref{eq:2} as a function of the temperature $T$
($\beta = 1/(k_B T)$) and of the chemical potentials $\mu_1^*$
and $\mu_2^*$ of the two species. Using the standard rules of
Statistical Mechanics, this will lead directly to the pressure $P =
k_B T \ln(\Xi) / V$ and to the mean occupation numbers $n_1^*$ and
$n_2^*$, and hence to the composition of the mixture. Phase separation
will be signalled by a discontinuous change of the mean occupation
number for well-defined values of $P$, $\mu_1^*$, and $\mu_2^*$. We
shall first consider the simplest version of the model which only
involves on-site interactions (i.e.\ $\Tens{\eta} = \Tens{0}$), before
examining the case including nearest neighbour interactions. 

\section{The case of on-site interactions only}
\label{sec:case-site-inter}

In the absence of couplings between particles on different sites, the
grand partition function \eqref{eq:2}, with $\Tens{\eta} = \Tens{0}$,
factorises into $L$ identical single site functions
$\Xi_{\text{site}}$: 
\begin{equation}
  \label{eq:3}
  \Xi = \left[\sum_{n_1,n_2 = 0}^{\infty} \frac{\exp(\beta
      \Vect{\mu}^* \cdot \Vect{n})}{n_1! n_2!} \exp\left(- \frac{1}{2}
      \Trans{\Vect{n}} \cdot \Tens{\epsilon} \cdot
      \Vect{n}\right)\right]^L
  = \Xi_{\text{site}}^L.
\end{equation}
The single site partition sum is not tractable analytically. For
sufficiently large chemical potentials, implying mean occupation
numbers $n_1^*, n_2^* \gg 1$, the latter may be treated as continuous
variables, and the sums in eq.~\eqref{eq:3} replaced by integrals:
\begin{equation}
  \label{eq:4}
  \Xi_{\text{site}} \approx \int_{n_1 = 0}^{\infty} \int_{n_2 =
    0}^{\infty} \frac{\exp\left(\beta \Vect{\mu}^* \cdot
      \Vect{n}-\frac{1}{2} \Trans{\Vect{n}}\cdot \Tens{\epsilon} \cdot
      \Vect{n}\right)}{\Gamma(n_1+1) \Gamma(n_2+1)} dn_1 dn_2.
\end{equation}
The integrand is sharply peaked around the most probable occupation
numbers $\Vect{n}^* = (n_1^*,n_2^*)$. Adopting the standard steepest
descent method, we expand the logarithm of the integrand around
$\Vect{n}^*$ to second order in $\delta \Vect{n}=
\Vect{n}-\Vect{n}^*$, i.e.:
\begin{multline}
  \label{eq:5}
  \beta \Vect{\mu}^*\cdot \Vect{n} -\frac{1}{2} \beta \Trans{\Vect{n}} \cdot
  \Tens{\epsilon} \cdot \Vect{n} - \sum_{\alpha = 1,2} \ln
  \Gamma(n_\alpha+1)\\
  \approx \beta  \Vect{\mu^*} \cdot \Vect{n^*} - \frac{1}{2} \beta \Trans{\Vect{n^*}} \cdot \Tens{\epsilon} \cdot \Vect{n^*} - \sum_{\alpha} \ln \Gamma(n^*_\alpha+1)\\
  +  \left(\beta \Vect{\mu^*} - \beta   \Trans{\Vect{n^*}} \cdot \Tens{\epsilon} -
    \begin{pmatrix}
      \psi(n^*_1 + 1)\\
      \psi(n^*_2 + 1)
    \end{pmatrix}
  \right) \cdot \delta \Vect{n}
  -  \frac{1}{2} \Trans{\delta \Vect{n}} \cdot \Tens{\sigma}
  \cdot \delta \Vect{n},
\end{multline}
where $\psi(n) = \text{d}\ln \Gamma(n)/\text{d}n$ is the digamma function,
and the covariance matrix $\Tens{\sigma}$ is given by:
\begin{equation}
  \label{eq:6}
  \Tens{\sigma}= 
  \begin{pmatrix}
    \beta \epsilon_{11} + \psi^{(1)}(n_1^*+1) & \beta\epsilon_{12}\\
    \beta \epsilon_{12} & \beta \epsilon_{22} + \psi^{(1)}(n_2^*+1)
  \end{pmatrix},
\end{equation}
with $\psi^{(1)}(n) = \text{d}\psi(n)/\text{d}n$ (trigamma function).
The location of the maximum follows from the extremum condition 
\begin{equation}
  \label{eq:7}
  \beta \Vect{\mu}^* - \beta \Trans{\Vect{n}^*} \cdot \Tens{\epsilon}
  - \begin{pmatrix}
    \psi(n^*_1 + 1)\\
    \psi(n^*_2 + 1)
  \end{pmatrix} = \Vect{0},
\end{equation}
which provides the relation between $\Vect{\mu}^*$ and $\Vect{n}^*$.
Inserting the expansion \eqref{eq:5} into the integrand in
eq.~\eqref{eq:4}, and extending the integral to negative occupation
numbers (which is again justified provided $n_\alpha^* \gg 1$), the
resulting Gaussian integral is easily evaluated and leads directly to
the following equation of state:
\begin{equation}
  \label{eq:8}
  \beta P v_0 = n_1^* + n_2^* + \frac{1}{2} \Trans{\Vect{n}^*}\cdot
  \Tens{\epsilon} \cdot \Vect{n}^* + \ln \left(\frac{2
  \pi}{\sqrt{\Det{\Tens{\sigma}}}}\right),
\end{equation}
where $v_0 = V/L$ is the volume per site.

The virial pressure in eq.~\eqref{eq:8}, derived on the basis of
steepest descent, is thermodynamically inconsistent with the relation
\eqref{eq:7} between chemical potentials and mean occupation numbers
(identical to their most probable values). If the latter is
integrated, an expression for the pressure follows which coincides
with eq.~\eqref{eq:8}, but without the logarithmic fluctuation term. The
situation is reminiscent of the RPA treatment of continuous penetrable
(e.g.\ Gaussian) core mixtures, where the virial and compressibility
routes lead to equations of state similar in structure to
eq.~\eqref{eq:8} with (virial) and without (compressibility) the last
term \cite{ref:5,ref:6,ref:7,ref:8}. The two equations of state become
identical in the high density limit pointing to the asymptotic
``mean-field fluid'' nature of systems of penetrable particles.

In view of the above analogy, it is not surprising that the equation
of state \eqref{eq:8} leads to phase separation between sites with
different mean occupation numbers when $\epsilon_{12}^2 > \epsilon_{11}
\epsilon_{22}$. This predicted phase separation is obviously spurious,
because of the effectively 0-dimensional nature of the model when
restriction is made to on-site couplings. The above steepest descent
treatment has two obvious shortcomings. The first deficiency is the
replacement of the discrete sums in eq.~\eqref{eq:3} by an integral,
as well as the extension of the lower bounds in the integral
\eqref{eq:4} to $-\infty$; these approximations are expected to break
down for low mean occupation numbers $n_\alpha^*$, or equivalently for
low chemical potentials $\mu_\alpha^*$. This deficiency can be easily
removed by reverting from the integrals to discrete sums over an
interval of values of $\Vect{n}^*$ in the vicinity of the most
probable values $\Vect{n}^*$, and retaining the continuous integrals
outside this interval. In the one-component case, the resulting
single-site partition function reads:
\begin{multline}
  \label{eq:9} 
  \Xi_{\text{site}} \approx
  \sum_{n=n_{\text{low}}}^{n_{\text{high}}-1} \frac{1}{n!}
  e^{\beta\mu^* n} e^{-\frac{1}{2} \beta \epsilon n^2} +
  \frac{\exp\{n^* \psi(n^*+1)\}}{\Gamma(n^*+1)} \exp\left\{\frac{1}{2}
    \beta\epsilon {n^*}^2\right\}
  \sqrt{\frac{\pi}{2\sigma}}\\
    \times\left[\erf\left(\sqrt{\frac{\sigma}{2}}(n_{\text{low}}-n^*)\right)+\erf\left(\sqrt{\frac{\sigma}{2}}n^*\right)-\erf\left(\sqrt{\frac{\sigma}{2}}(n_{\text{high}}-n^*)\right) + 1\right].
\end{multline}
where $n_{\text{low}} = \max(\ceil{n^*}-m,0)$ and $n_{\text{high}}-1 =
\ceil{n^*}+m -1$, and the ceil function $\ceil{}$ rounds to the
nearest higher integer.

\begin{figure}[htbp]
  \centering
  \includegraphics[angle=-90,width=0.9\textwidth]{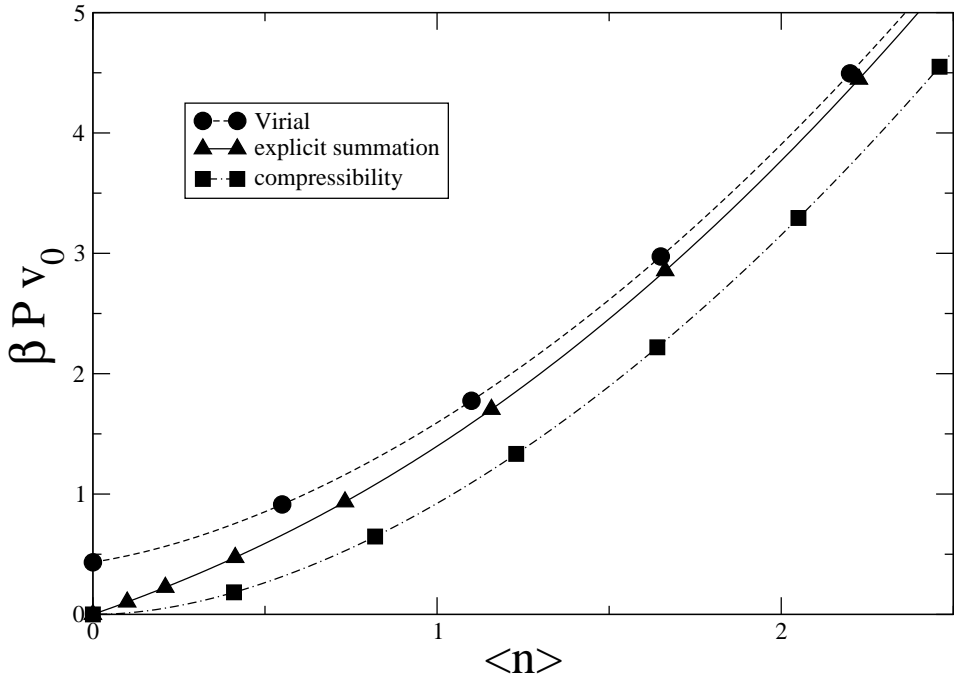}
  \caption{Equation of state of the one-component lattice gas with on
    site interaction energy $\epsilon = 1 k_B T$.  The normalised
    pressure $\beta P v_0$ is plotted as a function of the mean
    occupation number $\langle n \rangle$ of one cell. Shown are
    values obtained by using the steepest descent method (dashed
    curve, circles), the pressure consistent with the chemical
    potential \eqref{eq:7} (dash-dotted line, squares), and the
    pressure obtained from the grand canonical partition sum by
    explicitly summing $2m$ terms around the maximum of the Boltzmann
    factor as explained in the text with $m = 100$ (continuous line,
    triangles).  At this value the correction term in equation
    \eqref{eq:9} is well below $10^{-5}$, so that the resulting
    pressure can be considered nearly exact. The steepest descent
    method is seen to be very accurate except, as expected, at low
    densities.}
  \label{fig:1}
\end{figure}

The steepest descent result \eqref{eq:8} is recovered for $m= 0$ and
sufficiently high $n^*$, such that $\erf(n^*\sqrt{\sigma/2}) = 1$.
Results for the equation of state are shown in Fig.~\ref{fig:1}.
Eq.~\eqref{eq:8} is seen to yield surprisingly accurate results, even
at relatively low densities ($n^*\approx 1$), and only breaks down in
the $n^* \rightarrow 0$ limit.

\begin{figure}[htbp]
  \centering
  \includegraphics[width=0.9\textwidth]{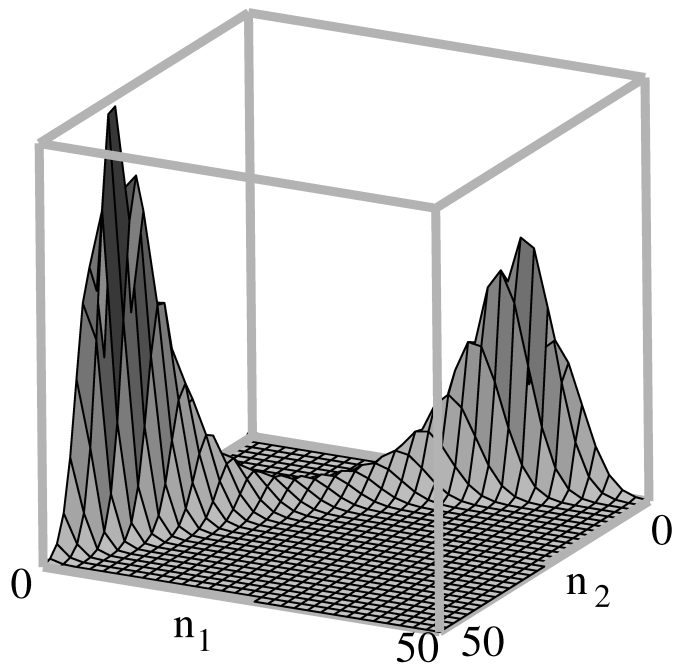}
  \caption{Boltzmann factor $\exp(\beta \Vect{\mu} \cdot \Vect{n} -  \beta E(\Vect{n}))/(n_1! n_2!)$ as a function of the mean occupation
  numbers $n_1, n_2$ for the lattice system with on site interaction
  $\epsilon_{11} = \epsilon_{22} = 0.1 k_B T$, $\epsilon_{12} = 0.2
  k_B T$ and no nearest neighbour interaction, at the chemical
  potentials $\mu_1 = 6.5 k_B T$, $\mu_2 = 6.0 k_B T$. The appearance
  of a secondary maximum is clearly seen.}
  \label{fig:2}
\end{figure}

\begin{figure}[htbp]
  \centering
  \includegraphics[angle=-90,width=0.9\textwidth]{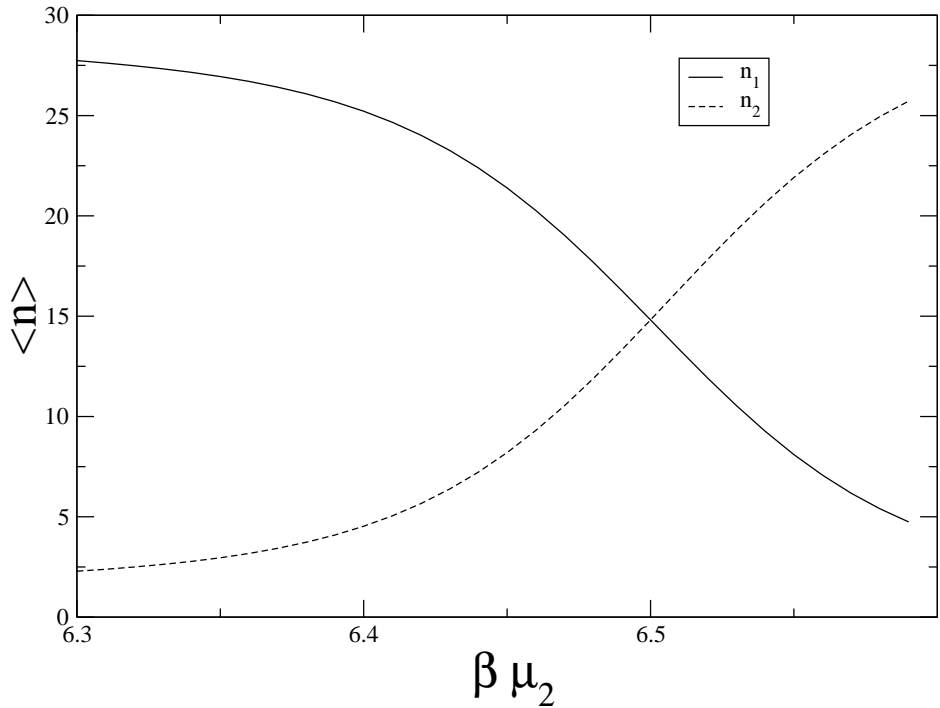}
  \caption{Mean occupation numbers $\langle n_1 \rangle$, $\langle n_2
    \rangle$ in a symmetric binary lattice gas mixture for on site
    interaction energies $\epsilon_{11} = \epsilon_{22} = 0.1 k_B T$
    $\epsilon_{12} = 0.2 k_B T$ and vanishing nearest neighbour
    interaction as a function of chemical potential $\mu_2$, keeping
    $\mu_1 = 6.5 k_B T$ fixed. The values were calculated using a
    simple generalization of equation \eqref{eq:9} As expected, the occupation numbers
    vary continuously with chemical potentials in this effectively
    $0$-dimensional system.}
  \label{fig:3}
\end{figure}

A more fundamental shortcoming of the steepest descent method is that
eq.~\eqref{eq:7} in fact predicts two maxima $\Vect{n}^*$, separated by a
saddle point. The situation is pictured in Fig.~\ref{fig:2}. The
steepest descent method includes only the contribution of the highest
(global) maximum of the integrand to the partition
function~\eqref{eq:4}. This approximation leads to a discontinuous
jump of the composition vector $\Vect{n}^*$ when the two maxima become
equal, and hence to a first order phase transition. In reality, when
the contributions of the two maxima of the configuration
integral~\eqref{eq:4} are properly included, the mean occupation
numbers $\langle n_1 \rangle$ and $\langle n_2\rangle$ vary
continuously with the chemical potentials of the two species, as
illustrated in Fig.~\ref{fig:3}, according to expectation.

\section{Including nearest neighbour interactions}
\label{sec:incl-near-neighb}

We now consider the model defined by eqns.~\eqref{eq:1} and
\eqref{eq:2} and including n.n.\ interactions between particles of
both species, i.e.\ $\Tens{\eta} \neq \Tens{0}$. The partition
function~\eqref{eq:2} no longer factorises into single site sums.
Proceeding as in Section~\ref{sec:case-site-inter}, we replace the
discrete sums over occupation numbers by integrals and approximate the
integrand by Gaussians centred on each of the local maxima. The
latter are determined by a generalisation of eq.~\eqref{eq:7}, namely

\begin{equation}
\label{eq:10}
\beta \Vect{\mu^*} = \beta \Trans{\Vect{n}^*_i} \cdot
  \Tens{\epsilon} + \sum_{i\text{n.n.} j} \beta \Trans{\Vect{n}^*_j} \cdot \Tens{\eta} +
  \begin{pmatrix}
  \psi(n^*_{1,i} + 1)\\
  \psi(n^*_{2,i} + 1)
  \end{pmatrix}; \quad 1 \le i \le L.
\end{equation}
In the limit $\Tens{\eta} \rightarrow \Tens{0}$, the $\Vect{n}^*_i$ on
different sites are independent and can each take two values,
corresponding to the two maxima found in
Section~\ref{sec:case-site-inter}. When the n.n.\ interactions are
switched on, the $\Vect{n}^*_i$ are coupled so that a complex energy
landscape emerges in $2L$-dimensional occupation-number space,
featuring a rapidly increasing number of maxima of the Boltzmann
factor.  By continuity one can
still expect $2^L$ solutions to eqns.~\eqref{eq:10} in the most
general case, at least for sufficiently weak coupling. The approximate grand partition function now reads
\begin{multline}
\label{eq:11}
  \Xi = \sum_{\{\Vect{n}^*_i\}} \exp\left[\frac{1}{2} \sum_i \Vect{n}_i^* \cdot \Tens{\epsilon} \cdot
  \Vect{n}_i^* + \sum_{i,\alpha} n_{i,\alpha}^* \psi(n_{i,\alpha}^*+1)
  - \sum_{i,\alpha} \ln \Gamma(n_{i,\alpha}^*+1) + 
  \sum_{\langle i j \rangle} \Vect{n}_i^* \cdot \Tens{\eta} \cdot
  \Vect{n}_j^*\right] \\
 \times \int d\delta\Vect{n}_1 \ldots \int d \delta \Vect{n}_L \exp\left(-  \left[\frac{1}{2} \sum_i \delta \Vect{n}_i \cdot \Tens{\sigma}_i
  \cdot \delta \Vect{n}_i  + \sum_{\langle ij \rangle} \beta
  \delta \Vect{n}_i \cdot \Tens{\eta} \cdot \delta \Vect{n}_j \right]  
\right)
\end{multline}
where the covariance matrices $\Tens{\sigma}_i$ are obvious
generalisations of eq.~\eqref{eq:6}, with $n_\alpha^*$ replaced by
$n^*_{\alpha,i}$. 

The grand partition function is thus expressed as a sum over all
``configurations'' $\{\Vect{n}_i^*\}$ which maximise the integrand in
the continuous representation generalising eq.~\eqref{eq:4}, i.e.\
which satisfy the set of eqns.~\eqref{eq:10}. The Gaussian fluctuation
integrals in the second factor depend on the configuration
$\{\Vect{n}_i^*\}$ through the covariance matrices $\Tens{\sigma}_i$.
For given $\{\Tens{\sigma}_i\}$ the coupled Gaussian integrals can be
calculated explicitly by a standard normal mode analysis.

The locations of the local maxima $\Vect{n}_i^*$ depend on
the occupation numbers of the neighbouring cells $\Vect{n}_j^*$ via
\eqref{eq:10}. Were that not the case, then the partition function
\eqref{eq:11} would exactly factor into a product of a partition sum
that is isomorphic to that of an Ising model (first factor in
\eqref{eq:11}) and the partition sum of a Gaussian model. The phase
behaviour of a very similar system has been investigated in Ref.
\cite{ref:8a}. The model we examine can thus only be mapped onto an
Ising (plus Gaussian) model in the limit of weak nearest neighbour
interactions $\Tens{\eta}$.

In order to gain further insight, we consider the ``ground state'',
corresponding to the low temperature or strong coupling limit. In that
limit all $\Vect{n}^*_i$ corresponding to the lowest minimum on the
total energy surface are expected to be equal ($\Vect{n}_i^* =
\Vect{n}^*$), and determined by a single extremum condition:
\begin{equation}
  \label{eq:12}
  \beta \Vect{\mu}^* = \beta \Trans{\Vect{n}^*} \cdot
  (\Tens{\epsilon}+q \Tens{\eta}) + 
  \begin{pmatrix}
  \psi(n^*_{1} + 1)\\
  \psi(n^*_{2} + 1)
  \end{pmatrix},
\end{equation}
where $q$ is the coordination number of the lattice. 
In the low temperature limit only the contribution of this one
configuration to the sum in equation \eqref{eq:11} is taken into
account. The second factor in this equation then becomes invariant
with respect to translations, so that the corresponding integral is
best evaluated in Fourier space. 

\begin{figure}[htbp]
  \centering
  \includegraphics[angle=-90,width=\textwidth]{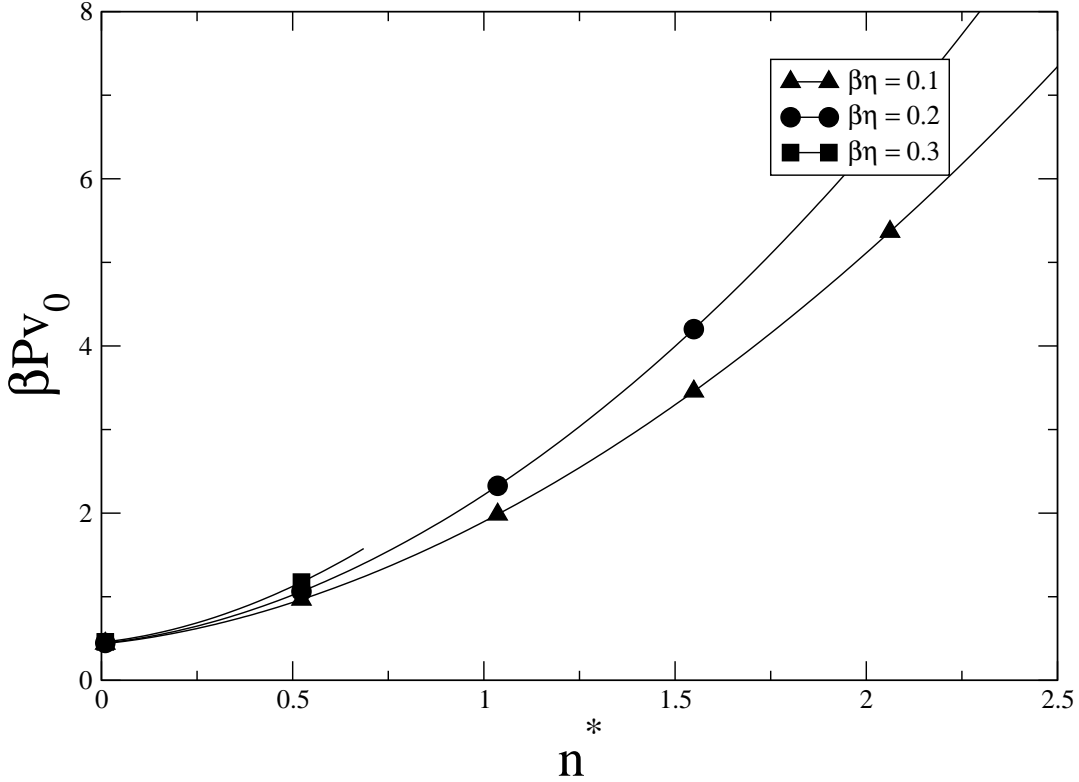}
  \caption{Virial equation of state of a one component lattice system with on
    site interaction $\epsilon = 1 k_B T$ and nearest neighbour
    interaction $\eta = 0.1 k_B T$ (triangles), $\eta = 0.2 k_B T$
    (circles), and $\eta = 0.3 k_B T$ (squares). For $\beta\eta > 1/6$
    our analysis predicts an instability of the homogeneous phase
    towards microphase separation beyond a certain density. For $\eta=0.1 k_B T$ this
    criterion is not obeyed and the fluid remains stable at all
    densities. For $\eta=0.2 k_B T$ the instability is at $n^* = 4.48$
    (outside the density range of this plot), and for $\eta = 0.3$ it is at $n^* = 0.69$.}
    \label{fig:4}
\end{figure}

\begin{figure}[htbp]
  \centering
  \includegraphics[angle=-90,width=\textwidth]{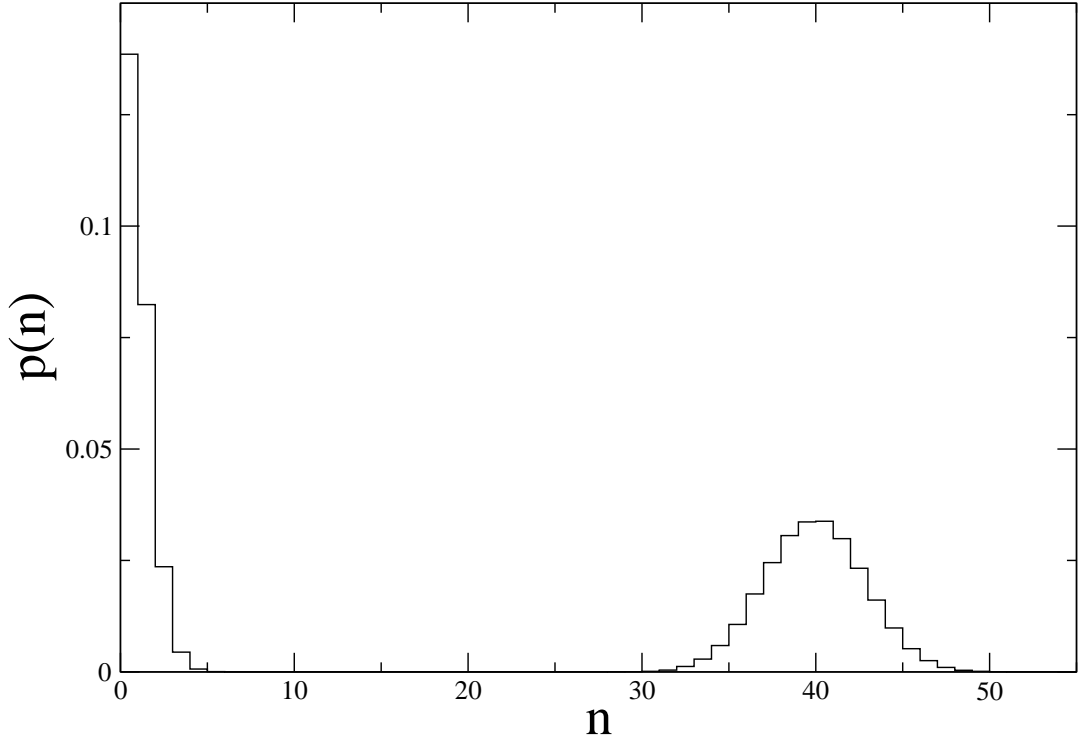}
  \caption{Histogram of the occupation numbers in a one component
    lattice system ($10 \times 10 \times 10$) with $\epsilon=0.1 k_B
    T$ and $\eta = 0.033 k_B T$, The probability $p(n)$ to find a
    lattice site with occupation number $n$ (averaged over 1~000~000
    steps after equilibration) is shown. The canonical simulation was
    started with 20 particles on each lattice site. Because of the
    large nearest neighbour interaction the initially homogeneous
    fluid develops into a microphase with a bimodal distribution of the
    occupation numbers.}
  \label{fig:4a}
\end{figure}

We therefore introduce the Fourier components of the density
fluctuations
\begin{equation}
  \label{eq:13}
  \delta \Vect{\tilde{n}}_{\Vect{k}} = \frac{1}{L} \sum_j \delta\Vect{n}_j e^{i
  \Vect{k} \cdot \Vect{R}_j},
\end{equation}
where $\Vect{r}_j$ denotes the position of the lattice site $j$ in
units of the lattice constant. Assuming Born-von Karman boundary
conditions the wave vector $\Vect{k}$ can only assume discrete values. 
The total interaction energy can be written in Fourier space as
\begin{equation}
  \label{eq:14}
  \frac{1}{2} \sum_i \delta \Vect{n}_i \cdot \Tens{\sigma}
  \cdot \delta \Vect{n}_i  + \sum_{\langle ij \rangle} \beta
  \delta \Vect{n}_i \cdot \Tens{\eta} \cdot \delta \Vect{n}_j
  = \frac{L}{2} \sum_{\Vect{k}} \delta
    \tilde{\Vect{n}}_{-\Vect{k}} \cdot \Tens{D}(\Vect{k}) \cdot \delta
  \tilde{\Vect{n}}_{\Vect{k}},
\end{equation}
where the matrix $\Tens{D}(\Vect{k})$ is given by  
\begin{equation}
  \label{eq:15}  
  \Tens{D}(\Vect{k}) =  \Tens{\sigma} +  \beta \Tens{\eta} \sum_{j \text{n.n.} 0}
  e^{i\Vect{k} \cdot \Vect{R}_j}. 
\end{equation}
For a simple cubic lattice we obtain
\begin{equation}
  \label{eq:16}
  \Tens{D}(\Vect{k}) = \Tens{\sigma} + 2 (\cos k_x + \cos k_y + \cos
  k_z) \beta \Tens{\eta}. 
\end{equation}
The Fourier modes of the density fluctuations thus decouple, so that
the multiple integral in equation
\eqref{eq:11} can be evaluated in Fourier space as a product of
Gaussian integrals
\begin{multline}
  \label{eq:17}
  \int d\delta\Vect{n}_1 \ldots \int d \delta \Vect{n}_L \exp\left(-
    \left[\frac{1}{2} \sum_i \delta \Vect{n}_i \cdot \Tens{\sigma}_i
      \cdot \delta \Vect{n}_i + \sum_{\langle ij \rangle} \beta \delta
      \Vect{n}_i \cdot \Tens{\eta} \cdot \delta \Vect{n}_j \right]
  \right)\\
  = L^L \int d\delta\tilde{\Vect{n}}_{\Vect{k}_1} \ldots
  \int d\delta\tilde{\Vect{n}}_{\Vect{k}_L} \exp \left(- \frac{L}{2}
    \sum_{\Vect{k}} \delta \tilde{\Vect{n}}_{\Vect{-k}}
    \cdot \Tens{D}(\Vect{k}) \cdot \delta \tilde{\Vect{n}}_{\Vect{k}},
  \right)
  = (2 \pi)^L \prod_{\Vect{k}} \frac{1}{\sqrt{\Det{\Tens{D}(\Vect{k})}}}
\end{multline}
The grand partition function then reduces to:
\begin{multline}
  \label{eq:18}
  \Xi = \left[\exp\left\{\frac{1}{2} \Vect{n}^* \cdot (\Tens{\epsilon}
      + q \Tens{\eta}) \cdot
      \Vect{n}^* + \sum_{\alpha} n_{\alpha}^* \psi(n_{i,\alpha}^*+1)
      -  \ln \Gamma(n_{\alpha}^*+1) \right\}\right]^L\\
   \times (2 \pi)^L \exp\left\{- \frac{L}{2} \int \frac{d\Vect{k}}{(2\pi)^3}
     \ln\left(\Det{\Tens{D}(\Vect{k})}\right)\right\},
\end{multline}
leading to the equation of state
\begin{equation}
  \label{eq:19}
  \beta P v_0 = \sum_\alpha [n_\alpha^* \psi(n_\alpha^*+1) - \ln
  \Gamma(n_\alpha^*+1)] + \frac{1}{2} \Vect{n}^* \cdot
  (\Tens{\epsilon}+ q \Tens{\eta}) \cdot \Vect{n}^* + \ln (2 \pi) -
 \frac{1}{2} \int \frac{d\Vect{k}}{(2 \pi)^3} \ln\left(\Det{\Tens{D}(\Vect{k})}\right).
\end{equation}
In the limit of large occupation numbers ($n_\alpha^* \gg 1$) and
vanishing nearest neighbour interaction ($\Tens{\eta} = \Tens{0}$) we
recover equation \eqref{eq:8}. As in the case without nearest
neighbour interaction, there is a thermodynamic inconsistency between
the e.o.s. \eqref{eq:19} and the dependence of the chemical potentials
on the occupation numbers \eqref{eq:12}. Integrating the latter would
again lead to an e.o.s. similar in structure to equation
\eqref{eq:19}, albeit without the final two logarithmic terms. The
virial equation of state is shown in Figure \ref{fig:4}. For
high nearest neighbour interaction ($\Tens{\epsilon} - q \Tens{\eta}$
no longer being positive definite) the fluctuation term diverges at $\Vect{k} =
(\pi, \pi, \pi)$. This instability indicates a transition of the ground state from
the homogeneous phase to a microfluid phase where high and low occupation
numbers alternate in a checkered fashion. To show that the  microphase exists
also for the one-component fluid (for $q \eta > \epsilon$) a canonical
Monte Carlo simulation was run on a $10\times 10 \times 10$ cubic
lattice ($q = 6$)
with the parameters $\epsilon = 0.1 k_B T$, $\eta = 0.033 k_B T$. Initially
every lattice site was filled with 20 particles. After equilibration a
histogram of the occupation numbers occuring over the next 1~000~000
steps was recorded. The result is shown in Figure \ref{fig:4a}. It is
clearly seen that the initial peak at $n = 20$  has vanished and a
bimodal distribiution around $n \approx 0$ and $n \approx 40$ has
developed. The development of the microphase is easy to understand
when we estimate and compare the energetic and entropic costs of the microphase
with (mean) occupation numbers $n_a$ and $n_b$ with those of a homogeneous
phase with average density $\bar{n} = (n_a + n_b) / 2$. The
entropy of the microphase per site is $s^{\text{micro}} = k_B [n_a \ln n_a + n_b \ln
n_b] / 2$, compared to that of the homogeneous phase $s^{\text{hom.}}
= k_B \bar{n} \ln \bar{n}$. The difference between the two is simply
the mixing entropy of the ideal gas, thereby favouring the homogeneous
density distribution. On the other hand the internal energy of the
microphase per site is $u^{\text{micro}} = \epsilon n_a^2 /4 + \epsilon
n_b^2 / 4 + q \eta n_a n_b / 2 $, compared to that of the
homogeneous phase $u^{\text{hom.}} = (\epsilon + q \eta) \bar{n}^2 /
2$. The difference between the two internal energies is
$u^{\text{micro}} - u^{\text{hom}} = (\epsilon - q \eta) (n_b - n_a)^2
/ 8$. For strong nearest neighbour interaction the microphase is
therefore energetically favoured. The nature of  the transition from
the homogeneous state to  the micro state will be investigated elsewhere.

\begin{figure}[htbp]
  \centering
  \includegraphics[angle=-90,width=\textwidth]{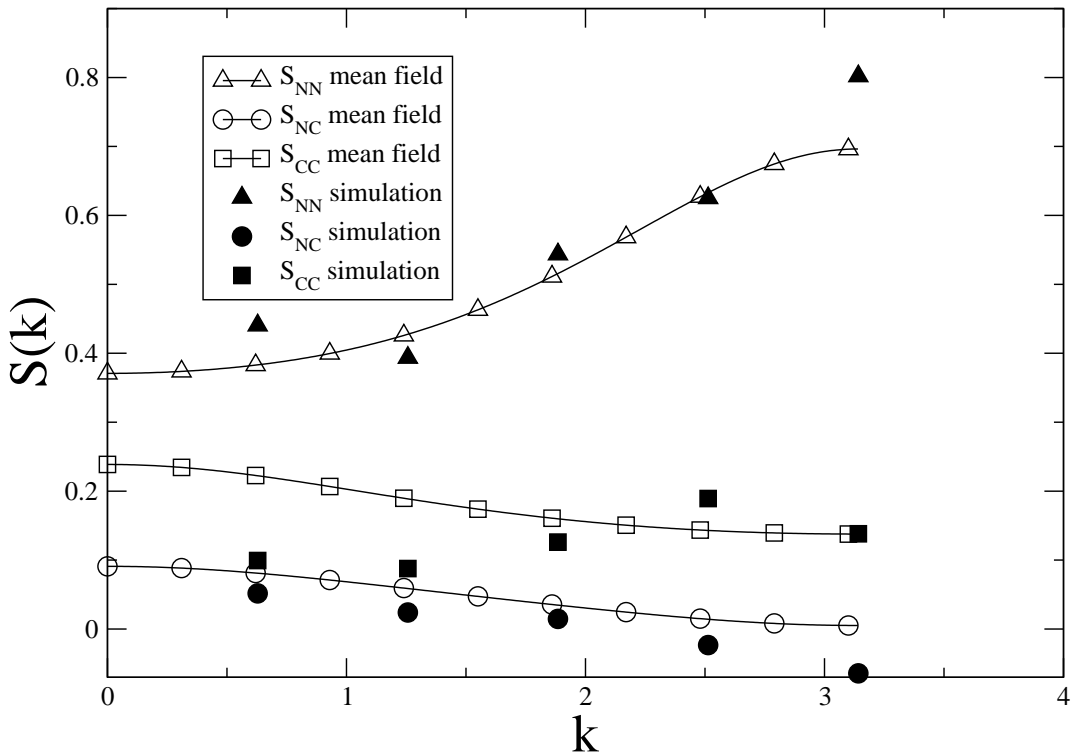}
  \caption{Bhatia-Thornton structure factors for the symmetric binary
    mixture with $\beta \epsilon_{11} = \beta\epsilon_{22} = 0.1$,
    $\beta\epsilon_{12} = 0.2$, $\Tens{\eta} = 0.1\Tens{\epsilon}$.
    The mean-field values (open symbols, lines) are compared to
    simulation data of a canonical Monte Carlo simulation (filled
    symbols) on a $10\times 10 \times 10$ lattice.  Shown are the
    density-density ($S_{nn}$, triangles), density-concentration
    ($S_{nc}$, circles) and concentration-concentration ($S_{cc}$,
    squares) structure factors along the $111$-direction for the
    occupation numbers $\langle n_1 \rangle = 10$, $\langle n_2
    \rangle = 1$. The statistics for the simulation data is known to
    be poor in $k$-space.}
  \label{fig:5}
\end{figure}

The explicit knowledge of the Fourier components of the density
fluctuations allows us to evaluate the structure factor of the fluid.
For $\Vect{k}\neq \Vect{0}$ we obtain
\begin{multline}
  \label{eq:20}
  S_{\alpha\beta}(\Vect{k}) =  
  \frac{1}{n^*_1 + n^*_2} \langle \tilde{\delta n}_\alpha(\Vect{k})
  \tilde{\delta n}_\beta(-\Vect{k}) \rangle\\
  = \frac{1}{n_1^* + n_2^*} \frac{\int d\tilde{\delta n}_1(\Vect{k})
    \int d \tilde{\delta n}_2(\Vect{k}) \tilde{\delta
      n}_\alpha(\Vect{k}) \tilde{\delta n}_\beta(\Vect{k})
    \exp(-\tilde{\Vect{\delta n}}(\Vect{k})\cdot \Tens{D}(\Vect{k})\cdot
    \tilde{\Vect{\delta n}}(\Vect{k}))}{\int d\tilde{\delta n}_1(\Vect{k})
    \int d \tilde{\delta n}_2(\Vect{k}) 
    \exp(-\tilde{\Vect{\delta n}}(\Vect{k})\cdot \Tens{D}(\Vect{k})\cdot
    \tilde{\Vect{\delta n}}(\Vect{k}))}\\
  = \frac{1}{n_1^* + n_2^*} \left(\Tens{D}^{-1}\right)_{\alpha\beta}(\Vect{k}).
\end{multline}
From the particle-particle structure factors $S_{\alpha\beta}$ the
Bhatia-Thornton structure factors \cite{ref:10} can be extracted.
These are compared with data from a canonical Monte Carlo simulation
in Figure \ref{fig:5} for the phase point $\langle n_1\rangle = 10$,
$\langle n_2\rangle = 1$. The agreement is reasonable, although there
are still fairly large statistical errors, a problem commonly
encountered in direct calculations of $S(k)$ (as opposed to Fourier
transforming $g(r)$). The error bars on the simulation results (not
shown) are rather large.

\begin{figure}[htbp]
  \centering
  \psfrag{energy}{{\LARGE $\beta E +\sum_\alpha \ln (n_\alpha^*!) - \sum_\alpha
    \beta \mu_\alpha n_\alpha^*$}}
  \psfrag{lambda}{{\LARGE $\lambda$}}
  \psfrag{vanish}{\includegraphics[width=0.4\textwidth]{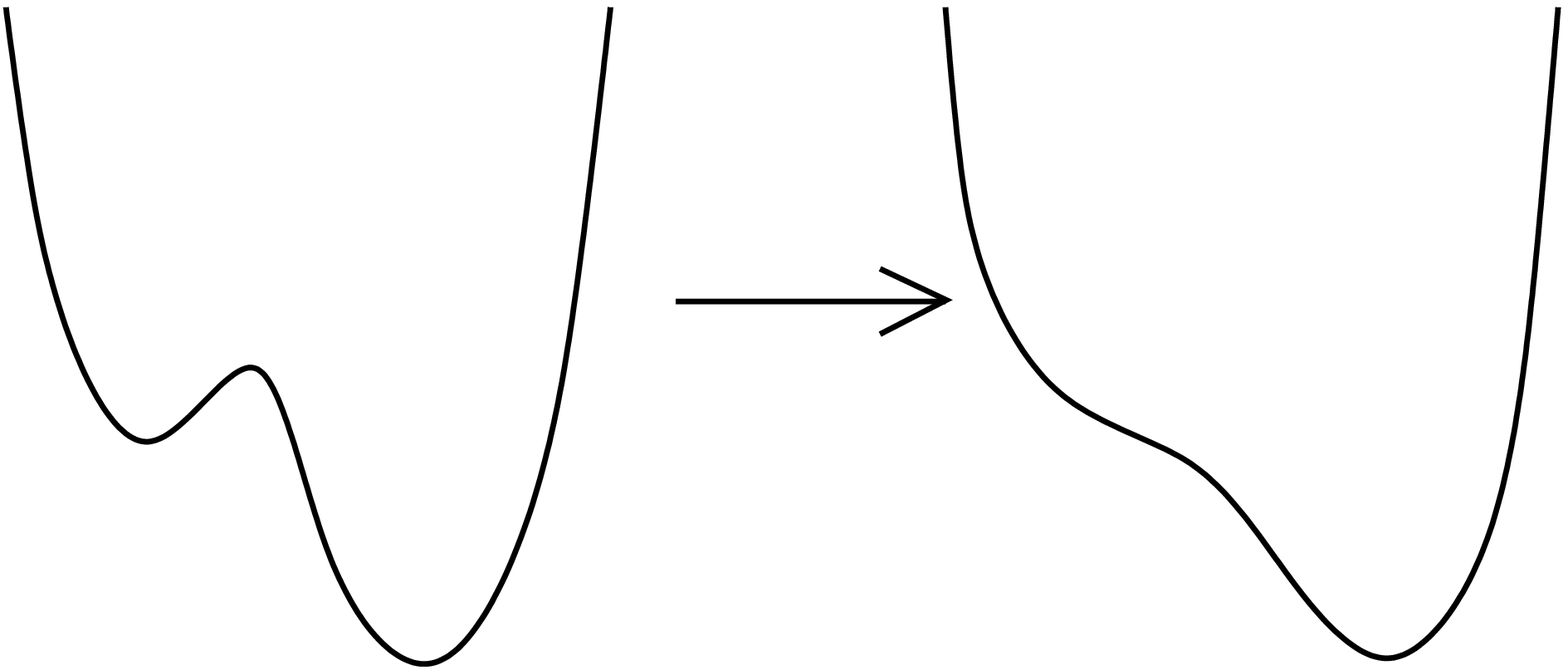}}
  \includegraphics[angle=-90,width=\textwidth]{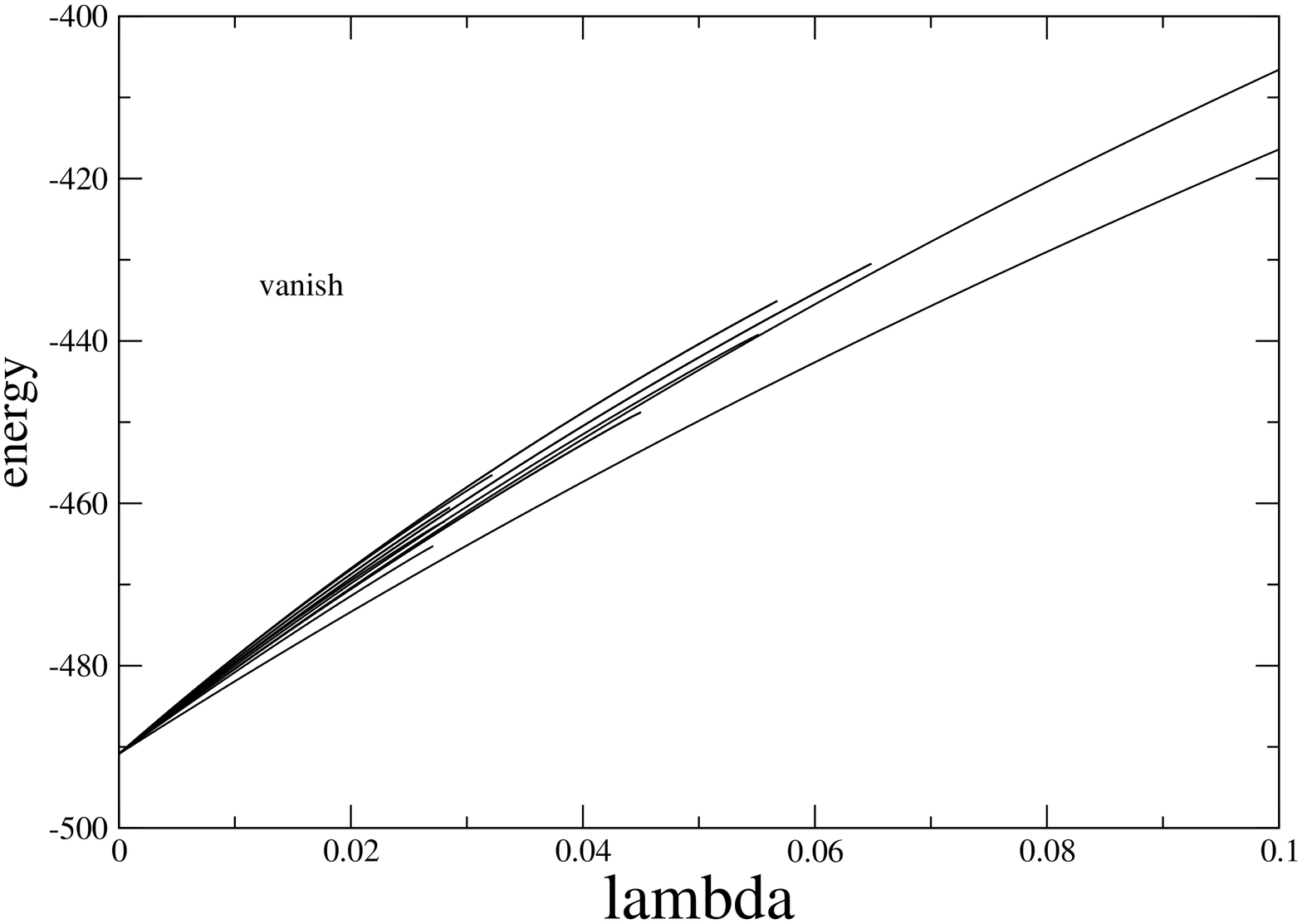} 
  \caption{Local energy minima for a $2 \times 2 \times 2$ lattice
  system at $\mu_1 = \mu_2 = 6.5 k_B T$. The energy minima are
  degenerate for vanishing nearest neighbour interaction ($\lambda =
  0$) and split for finite $\lambda$ into sets of local minima related by
  symmetry of the lattice. At certain critical $\lambda$, some minima
  turn into saddle points and vanish, as visualized in the inset.}
  \label{fig:6}
\end{figure}

\begin{figure}[htbp]
  \centering
  \psfrag{energies}{{\LARGE $\beta E +\sum_\alpha \ln (n_\alpha^*!) - \sum_\alpha
    \beta \mu_\alpha n_\alpha^*$}}
  \psfrag{lambda}{{\LARGE $\lambda$}}
  \includegraphics[angle=-90,width=\textwidth]{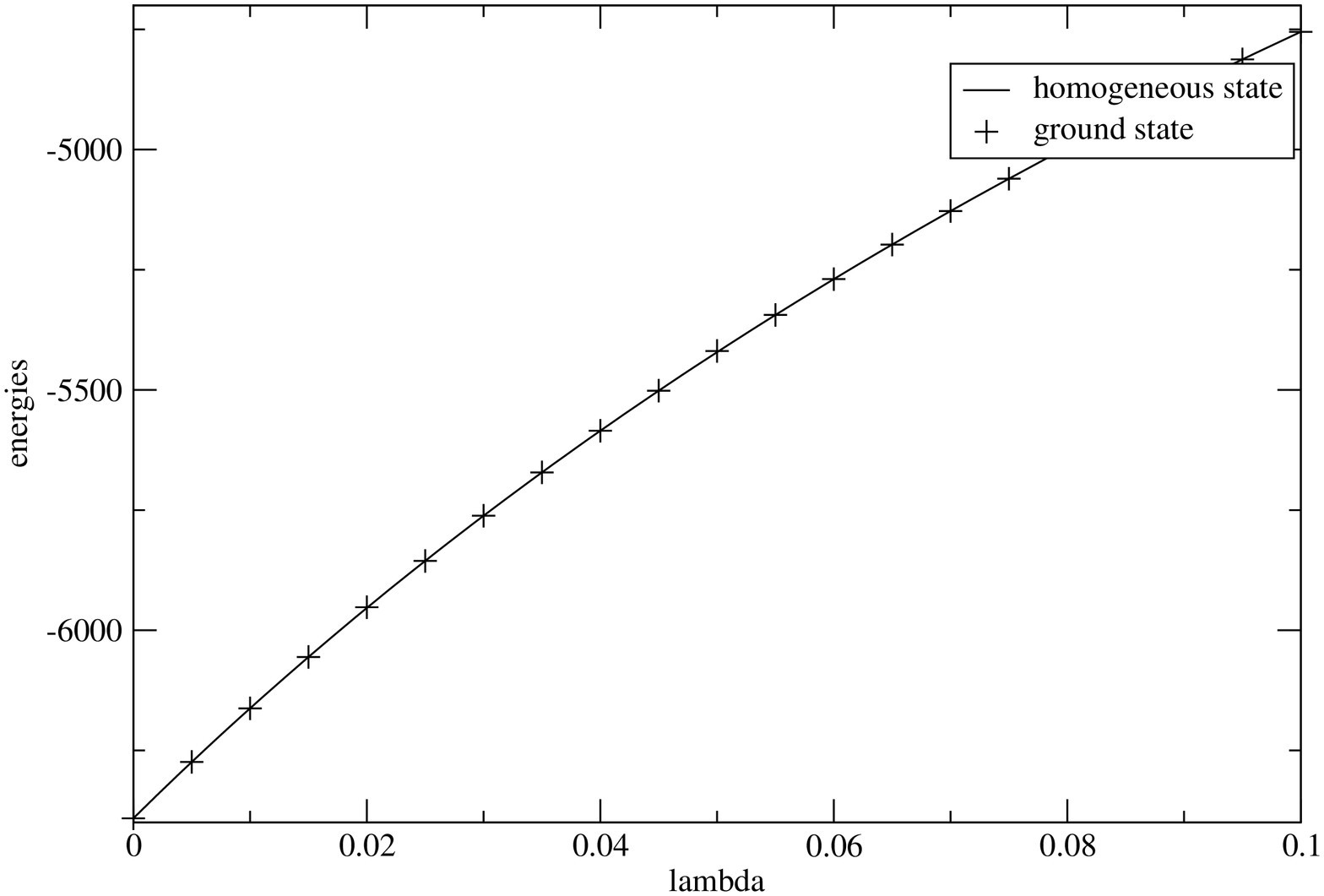}
  \caption{Absolute ground state energies in a $5\times 5 \times 5$
  lattice at $\beta \mu_1 = \beta \mu_2 = 6.5$ as found by the
  optimization procedure described in the text (plusses) compared to
  the energy of the homogeneous state, as a function of the nearest
  neighbour interaction $\Tens{\eta} = \lambda \Tens{\epsilon}$. On
  site interactions are $\epsilon_{11} = \epsilon_{22} = 0.1 k_B T$,
  $\epsilon_{12} = 0.2 k_B T$.}
  \label{fig:7}
\end{figure}

\begin{figure}[htbp]
  \centering
\psfrag{lambda}{{\LARGE $\lambda$}}
\psfrag{eground}{{\LARGE $\beta E - \beta E_0$}}
  \includegraphics[angle=-90,width=\textwidth]{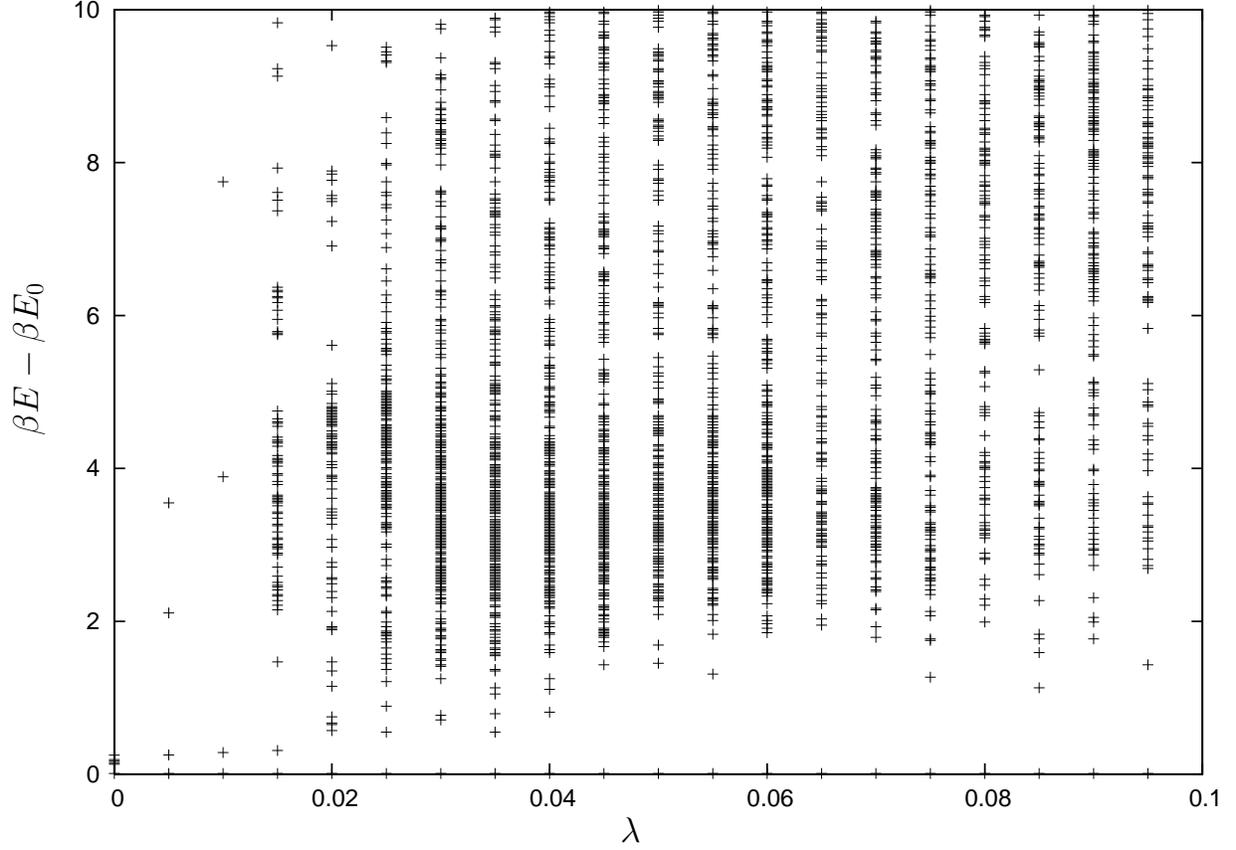}
  \caption{Local energy minima with respect to the ground state, as found by the optimization procedure
    described in the text. The on site interactions are $\epsilon_{11} =
    \epsilon_{22} = 0.1 k_B T$, $\epsilon_{12} = 0.2 k_B T$. The
    nearest neighbour interactions are $\Tens{\eta} = \lambda
    \Tens{\epsilon}$ with $\lambda = 0, 0.005, 0.01, \ldots, 0.1$. It
    is clearly seen that the energy gap from the ground state to the
    first excited state increases with increasing nearest neighbour
    interaction. The ground state approximation should therefore
    become more accurate for stronger nearest neighbour interaction.}
  \label{fig:8}
\end{figure}

In the calculations above we neglected all contributions from
configurations in the partition sum \eqref{eq:11} other than the
(presumed) ground state, where all occupation numbers are the same.
This is of course only justified when the other configurations have
much higher energies. To check this hypothesis, we explore the energy
landscape of small lattice systems with $\mu_1 = \mu_2$. For a $2
\times 2 \times 2$ lattice systems, one can explicitly enumerate all
local energy minima. In the case of no nearest neighbour interactions
the Boltzmann factor has $2^8$ degenerate local maxima. Slowly
switching on nearest neighbour interactions by setting $\Tens{\eta} =
\lambda \Tens{\epsilon}$ and varying $\lambda$, one can numerically
follow the local maxima of the Boltzmann factor. Due to the high
symmetry, many local minima stay degenerate even at finite nearest
neighbour interaction.  When increasing $\Tens{\eta}$, some local
minima turn into saddle points and vanish, as illustrated in Figure
\ref{fig:6}. The remaining excited energy states depending on
$\lambda$ are shown in the same figure. The increasing gap between the
ground state and the excited states is clearly seen.
For larger lattices the explicit enumeration of all local minima is no
longer feasible because of their large number. We therefore resort to
a simple Metropolis algorithm employed in numerical optimization
problems \cite{ref:9}. Instead of systematically searching the $2L$
dimensional occupation number state, we generate trial configurations
randomly from the previously found local minimum, analogous to the
random particle displacements used in Monte Carlo simulations. After each random step in the space of
occupation numbers a local minimization of the energy is performed.
The random step is then accepted or rejected based on the energy
difference of the two local minima. Effectively the very rugged energy
landscape is therefore replaced with a locally constant function,
thereby eliminating the numerical problems one usually encounters when
solving global optimization problems. In Figure \ref{fig:7} the
resulting local energy minima are shown together with the (presumed)
homogeneous ground state. All the minima lie on or slightly above the presumed
ground state, thereby supporting our hypothesis that the ground state
is the homogeneous state. In Figure \ref{fig:8} the difference between
the local minima and the ground state energy is plotted. Even
though some local minima are not found during the search, one can
clearly see a trend of an increasing gap between the ground state and
the first excited state when the nearest neighbour interaction
increases. For a nearest neighour interaction that equals 10 percent
of the on-site interaction one finds a gap of approximately $1.5 k_B T$.
Even stronger nearest neighbour interaction should further widen the
gap, leading to a higher accuracy of the ground state approximation.

\clearpage
\section{Gibbs ensemble Simulations}
\label{sec:simulations}

\begin{figure}[htbp]
  \centering
  \includegraphics[angle=-90,width=\textwidth]{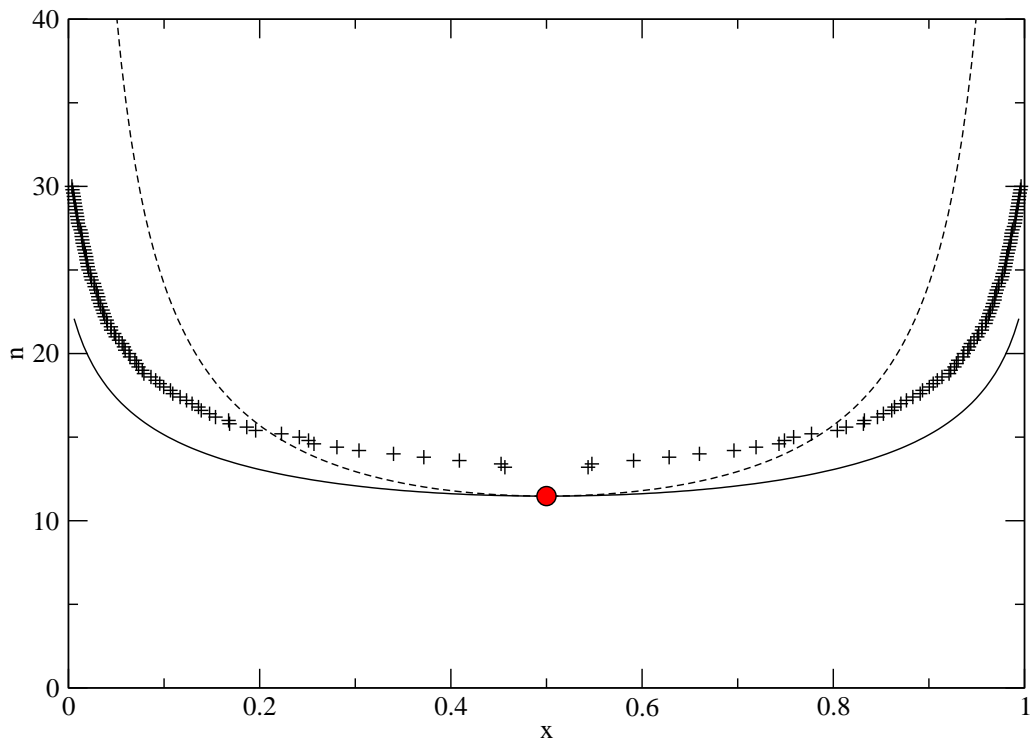}
  \caption{Phase diagram of a symmetric binary lattice system with $\epsilon_{11} =
    \epsilon_{22} = 0.1 k_B T$, $\epsilon_{12} = 0.2 k_B T$, $\eta =
    0.1 \epsilon$.  Shown are the phase coexistence points of the
    Gibbs ensemble simulation (crosses) and the predictions of the binodal
    (solid line) and spinodal (dashed line) based on the ground state
    approximation. No microphase separation is expected for such low
    nearest neighbour interactions.}
  \label{fig:9}
\end{figure}

\begin{figure}[htbp]
  \centering
  \includegraphics[angle=-90,width=\textwidth]{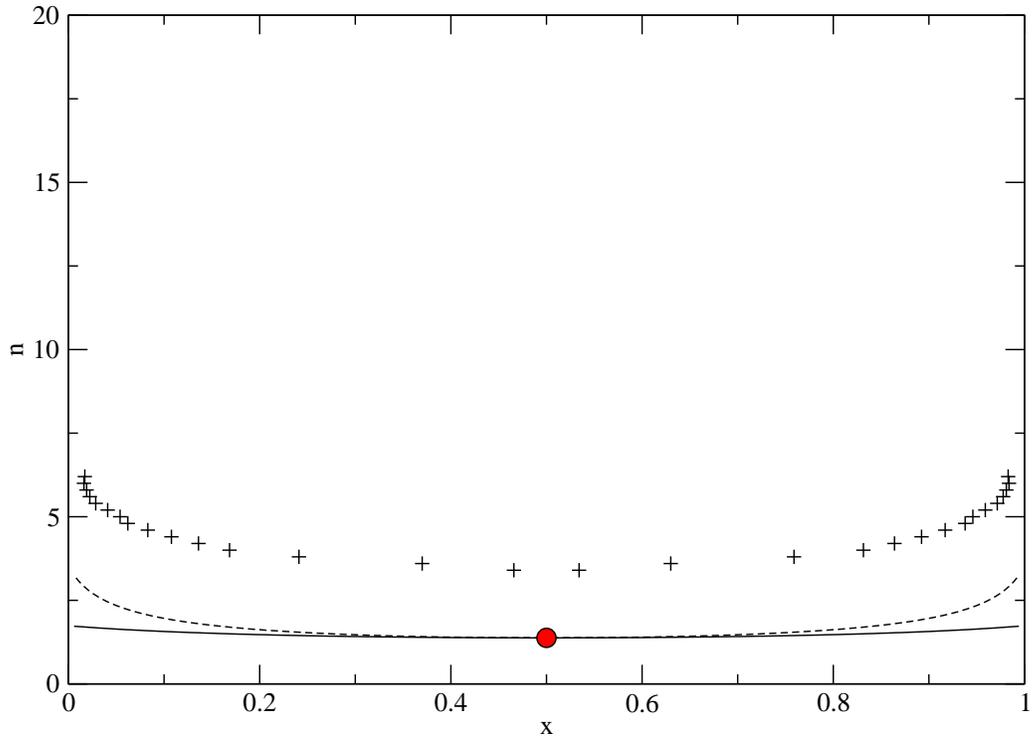}
  \caption{Phase diagram of a symmetric binary lattice system with 
    $\epsilon_{11} = \epsilon_{22} = 0.5 k_B T$, $\epsilon_{12} = 1 k_B T$,
    $\eta = 0.1 \epsilon$ as in Figure \ref{fig:9}.}
  \label{fig:10}
\end{figure}

To test the validity of the ground state approximation, Gibbs ensemble Monte Carlo
simulations \cite{ref:11} of the model have been performed on a $10\times 10\times
10$ lattice. After an equilibration time of 1~000~000 steps two lattices
were allowed to exchange particles and random moves within
each lattice were attempted according to the standard Metropolis algorithm. The simulations were restricted to symmetric
interactions with $\epsilon_{11} = \epsilon_{22}$ and $\eta_{11} =
\eta_{22}$. Since in these systems the equation of state must be
invariant with respect to relabelling of the species (or equivalently
mirroring the concentration $x \rightarrow 1-x$), volume changes of
the lattice were not necessary, i.\ e.\ the two lattice systems were
automatically under equal pressure. Two typical phase diagrams are shown in
Figure \ref{fig:9} and \ref{fig:10}. It is seen that the ground state
approximation predicts the correct shape of the phase diagram, albeit
underestimates the densities of the coexistencing phases. The discrepancy becomes
larger for larger interaction energies, when the contributions of the
excited states are no longer negligible, and the ground state
approximation becomes less accurate.

\clearpage
\section{Conclusion}

The lattice model introduced in this paper, which allows for multiple 
occupancy of lattice sites by one or two species, with positive energy 
penalties, and includes repulsive nearest neighbour interactions, leads 
to intersting phase behaviour. In the one-component version, a steepest 
descent analysis allowing for gaussian fluctuations, predicts microphase 
separation onto two interpenetrating sub-lattices with different mean 
occupation numbers ("checkered" phase) when the ratio $\eta / \epsilon$ of 
nearest-neighbour to on-site couplings exceeds a threshold. This prediction 
is confirmed by Monte Carlo simulations.
   
The two-component extension of the model leads to a macroscopic
demixing transition into phases of different concentrations of the two
species. The steepest descent analysis predicts such a phase
separation, even in the absence of nearest-neighbour interactions, but
this is shown to be due to the expected break-down of the
approximation. In the presence of nearest-neighbour interactions, the
steepest-descent analysis corresponds to a "ground state"
approximation. A numerical search of local minima of the energy
surface points to a significant gap ($>1 k_B T$) between the homogeneous
"ground-state" and the first "excited states", thus providing support
for the validity of the approximation. The predicted phase diagrams
agree reasonably well with the results of Gibbs ensemble MC
simulations. The present model is directly inspired by the continuous
"Gaussian core" model, first introduced by Stillinger \cite{ref:4},
and which has recently been shown to provide an adequate
coarse-grained description of interacting, interpenetrating polymer
coils \cite{ref:3}. The corresponding lattice model, considered here, appears to
have some unorthodox features. Contrary to more familiar lattice gas
models, it is not isomorphous to an Ising spin model. Hence it is not yet
clear whether its critical behaviour belongs to the Ising universality
class. The binary version is expected to exhibit both the microphase
separation found here in the one-component case, and the macroscopic
phase separation anologuous to that of the continuous binary gaussian
core model \cite{ref:6,ref:7,ref:8}. The interplay between these two
phase transitions should lead to interesting and novel behaviour,
which will be explored in future work.

\begin{acknowledgments}
\label{sec:acknoledgements}
RF would like to thank the Oppenheimer fund for financial support. AAL
wishes to acknowledge the Royal Society for their funding.
\end{acknowledgments}

\end{document}